\documentclass[twocolumn,aps,prl,showpacs,preprintnumbers,amsmath,amssymb, superscriptaddress]{revtex4-1}
\pdfoutput=1

\usepackage{bm}
\usepackage{graphicx}
\usepackage{color,xspace}

\renewcommand{\vec}{\mathbf}

\begin{document}

\title{Observation of two distinct $d_{xz}/d_{yz}$ band splittings in FeSe}

\author{P. Zhang}
\affiliation{Beijing National Laboratory for Condensed Matter Physics, and Institute of Physics, Chinese Academy of Sciences, Beijing 100190, China}
\author{T. Qian}
\affiliation{Beijing National Laboratory for Condensed Matter Physics, and Institute of Physics, Chinese Academy of Sciences, Beijing 100190, China}
\author{P. Richard}\email{p.richard@iphy.ac.cn}
\affiliation{Beijing National Laboratory for Condensed Matter Physics, and Institute of Physics, Chinese Academy of Sciences, Beijing 100190, China}
\affiliation{Collaborative Innovation Center of Quantum Matter, Beijing, China}
\author{X. P. Wang}
\affiliation{State Key Laboratory for Low-Dimensional Quantum Physics, Department of Physics, Tsinghua University, Beijing 100084,China}
\affiliation{Collaborative Innovation Center of Quantum Matter, Beijing, China}
\affiliation{Beijing National Laboratory for Condensed Matter Physics, and Institute of Physics, Chinese Academy of Sciences, Beijing 100190, China}
\author{H. Miao}
\affiliation{Beijing National Laboratory for Condensed Matter Physics, and Institute of Physics, Chinese Academy of Sciences, Beijing 100190, China}
\author{B. Q. Lv}
\affiliation{Beijing National Laboratory for Condensed Matter Physics, and Institute of Physics, Chinese Academy of Sciences, Beijing 100190, China}
\author{B. B. Fu}
\affiliation{Beijing National Laboratory for Condensed Matter Physics, and Institute of Physics, Chinese Academy of Sciences, Beijing 100190, China}
\author{T. Wolf}
\affiliation{Institut f\"{u}r Festk\"{o}rperphysik, Karlsruhe Institute for Technology, Karlsruhe 76021, Germany}
\author{C. Meingast}
\affiliation{Institut f\"{u}r Festk\"{o}rperphysik, Karlsruhe Institute for Technology, Karlsruhe 76021, Germany}
\author{X. X. Wu}
\affiliation{Beijing National Laboratory for Condensed Matter Physics, and Institute of Physics, Chinese Academy of Sciences, Beijing 100190, China}
\author{Z. Q. Wang}
\affiliation{Department of Physics, Boston College, Chestnut Hill, MA 02467, USA}
\affiliation{Beijing National Laboratory for Condensed Matter Physics, and Institute of Physics, Chinese Academy of Sciences, Beijing 100190, China}
\author{J. P. Hu}\email{jphu@iphy.ac.cn}
\affiliation{Beijing National Laboratory for Condensed Matter Physics, and Institute of Physics, Chinese Academy of Sciences, Beijing 100190, China}
\affiliation{Collaborative Innovation Center of Quantum Matter, Beijing, China}
\author{H. Ding}\email{dingh@iphy.ac.cn}
\affiliation{Beijing National Laboratory for Condensed Matter Physics, and Institute of Physics, Chinese Academy of Sciences, Beijing 100190, China}
\affiliation{Collaborative Innovation Center of Quantum Matter, Beijing, China}

\date{\today}

\begin{abstract}
We report the temperature evolution of the detailed electronic band structure in FeSe single-crystals measured by angle-resolved photoemission spectroscopy (ARPES), including the degeneracy removal of the $d_{xz}$ and $d_{yz}$ orbitals at the $\Gamma$/Z  and M points, and the orbital-selective hybridization between the $d_{xy}$ and $d_{xz/yz}$ orbitals. The temperature dependences of the splittings at the $\Gamma$/Z  and M points are different, indicating that they are controlled by different order parameters. The splitting at the M point is closely related to the structural transition and is attributed to orbital ordering defined on Fe-Fe bonds with a $d$-wave form in the reciprocal space that breaks the rotational symmetry. In contrast, the band splitting at the $\Gamma$ points remains at temperature far above the structural transition. Although the origin of this latter splitting remains unclear, our experimental results exclude the previously proposed ferro-orbital ordering scenario. 
\end{abstract}







\maketitle



Several experimental studies report the breakdown of the rotational symmetry in parent and underdoped compounds of Fe-based superconductors (FeSCs) \cite{FisherScience2010, ProzorovPRB2010, DegiorgiEPL2011, LuPNAS2011} that is commonly refereed to as nematicity. Its origin is highly debated since both magnetic \cite{KivelsonPRB2008, QiPRB2009, PaulPRB2010, DagottoPRL2013} and orbital \cite{KuPRL2009, OnariPRL2012, ZeyherPRB2013, LittlewoodPRB2013} fluctuations or orderings can lead to nematicity. Although strong support is given to magnetic-driven nematicity in iron-pncitides \cite{FernandesNP2014} where the orthorhombic lattice distortion is always accompanied by a collinear magnetic order at a temperature equal to or below the lattice transition temperature, this mechanism is questioned in FeSe, which exhibits an orthorhombic lattice distortion below the distortion transition temperature $T_s\sim 90$ K without any trace of magnetic order. As a direct signature of the electronic anisotropy between the $x$ and $y$ directions in the nematic state, previous angle-resolved photoemission spectroscopy (ARPES) studies \cite{TakahashiPRL2014, ShimojimaPRB2014, Coldeaarxiv2015} revealed a splitting between the otherwise degenerate Fe $3d_{xz}$ and Fe $3d_{yz}$ orbitals at the M point of the Brillouin zone (BZ). This splitting is widely believed to be a key evidence for ferro-orbital ordering in the nematic phase \cite{KuPRL2009,FengPRB2012,FernandesPRB2014,BuchnerNM2015}.

In this letter we report the existence of two distinct splittings between the $d_{xz}/d_{yz}$ bands of FeSe single-crystals near the Fermi level ($E_F$). We show that a first splitting decreases with temperature increasing, and disappears at about 100 - 120 K, which is slightly higher than $T_s$, suggesting that it is caused by short-range orbital order or fluctuations related to the structural transition. This splitting has a $d$-wave form breaking rotational symmetry \cite{LiJPCM2015} that is the largest at the M point and that is inconsistent with ferro-orbital ordering. In addition, we observe a splitting at the $\Gamma$ point that is rather insensitive to temperature up to 150 K, way above $T_s$. Due to the strong orbital-selectivity of the hybridization between the $d_{xy}$ band and the $d_{xz}$ and $d_{yz}$ orbitals, we conclude that the splitting at the $\Gamma$ point is not simply due to spin-orbit coupling (SOC).



High-quality single-crystals of $\beta$-FeSe were grown by the KCl/AlCl$_3$ chemical vapor transport method \cite{MeingastPRB2013}. The $T_c$ was determined to be 9 K from magnetization measurements and a structural transition is observed around 90 K. ARPES measurements were performed at the Dreamline beamline of the Shanghai Synchrotron Radiation Facility (SSRF) using a VG-Scienta D80 electron analyzer, and at the Institute of Physics, Chinese Academy of Sciences, using a R4000 analyzer and a helium discharge lamp. The angular resolution was set to 0.2$^\circ$. Clean surfaces for the ARPES measurements were obtained by cleaving the samples \emph{in situ} in a working vacuum better than $5\times 10^{-11}$ Torr. In the text, we label the momentum values with respect to the 1 Fe/unit cell BZ.

\begin{figure}[!t]
\begin{center}
\includegraphics[width=0.49\textwidth]{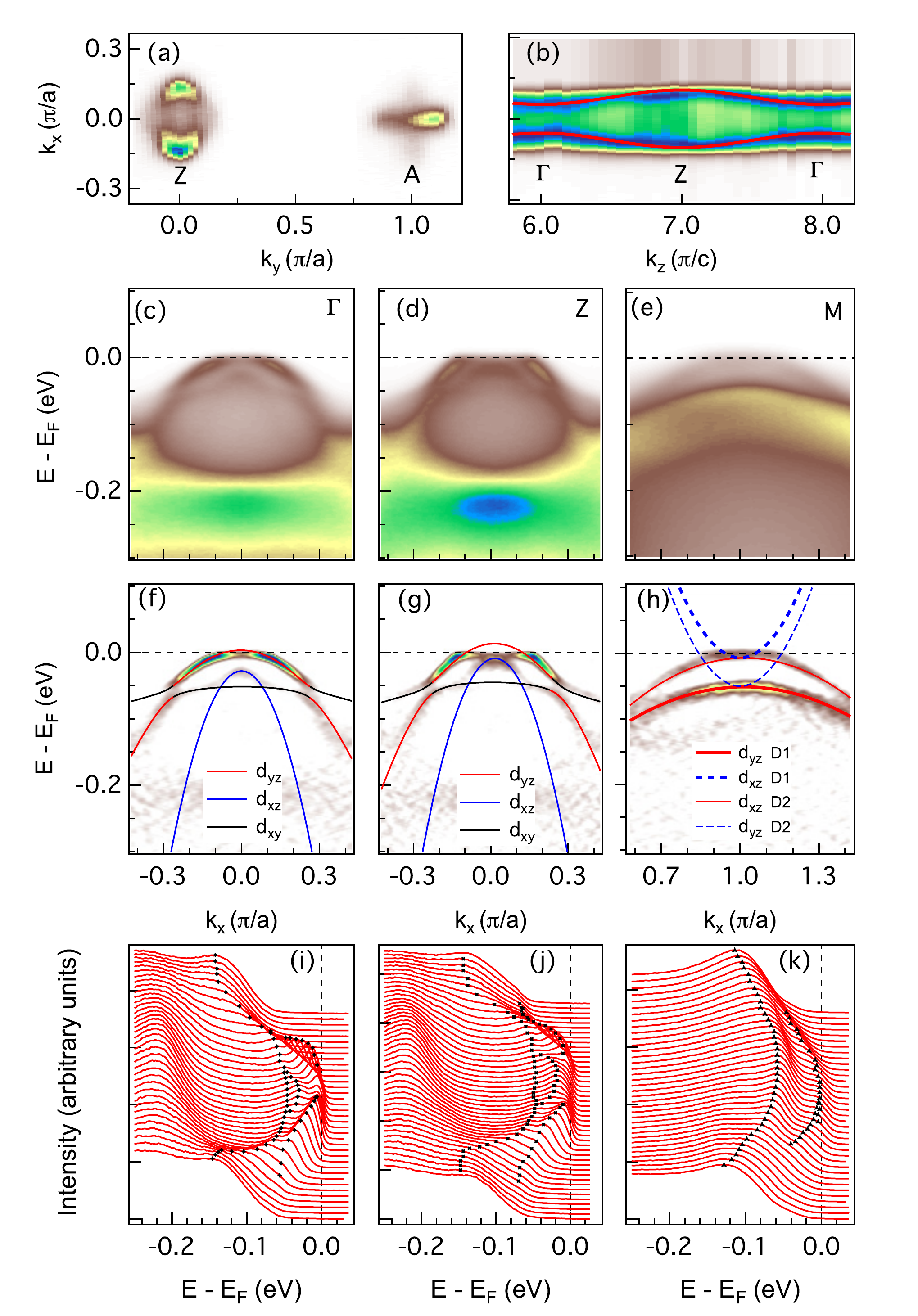}
\end{center}
 \caption{\label{band} (\emph{a}) FS mapping along the Z(0, 0, $\pi$)-A(0, $\pi$, $\pi$) direction, recorded with unpolarized He I$\alpha$ photons. (\emph{b}) FS mapping along the $\Gamma$ (0, 0, 0) - Z direction obtained in the $\sigma$ geometry. The red solid lines indicate the $k_z$ dispersion. (\emph{c - e}) Band structure at $\Gamma$, Z and M (0, $\pi$, 0) along $\Gamma$-M or Z-A with $T < T_s$, recorded in the $\sigma$ geometry. (\emph{f - h}) 2D curvature of (\emph{c - e}). The red, blue and black lines indicate the dispersions of the $d_{yz}$, $d_{xz}$ and $d_{xy}$ orbitals, respectively. The thick and thin lines in (\emph{h}) correspond to 2 different domains. (\emph{i - k}) EDC plots of (\emph{c - e}). The black dots indicate the EDC peaks.}
\end{figure}

\begin{figure*}[!htb]
\begin{center}
\includegraphics[width=\textwidth]{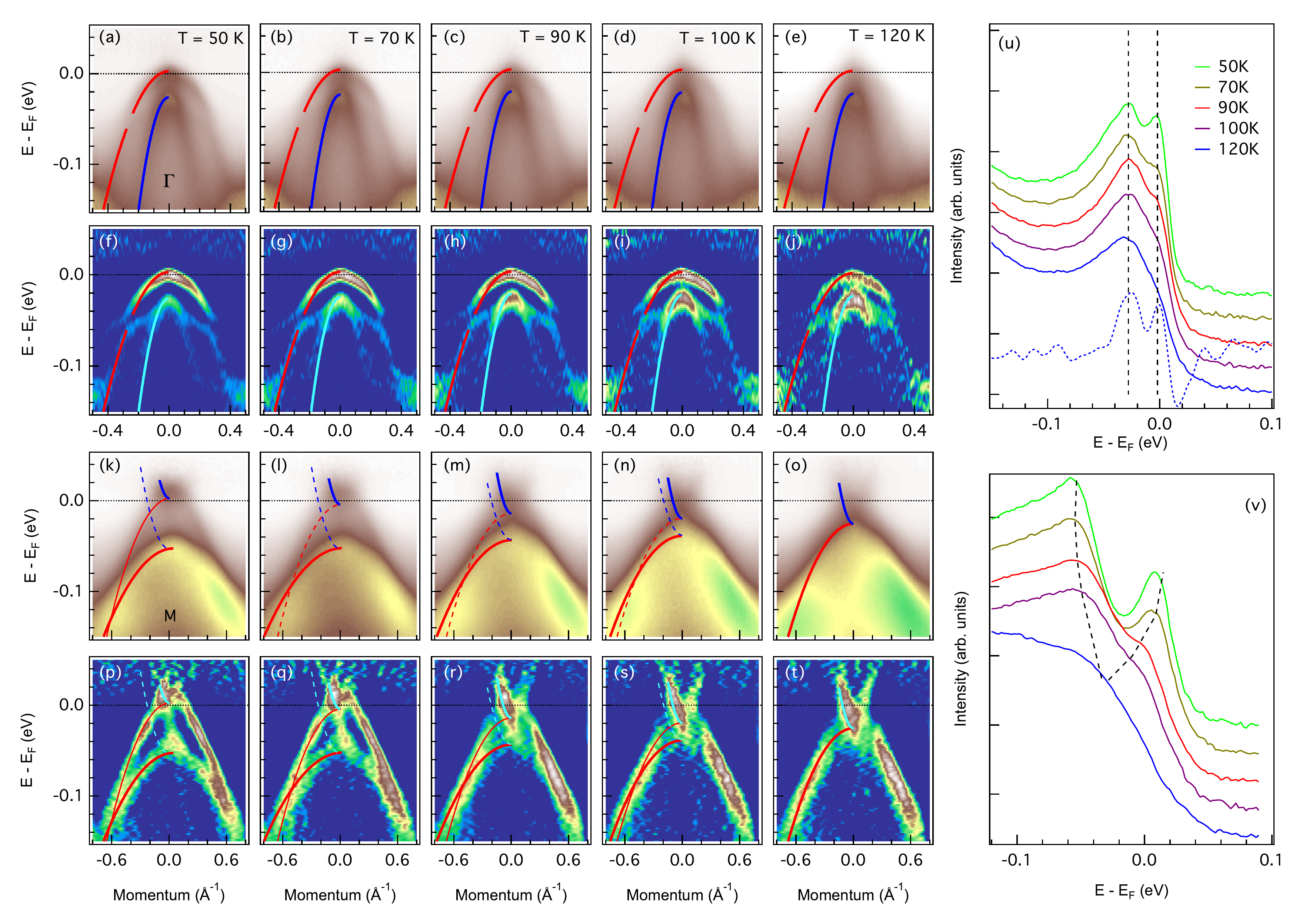}
\end{center}
 \caption{\label{split} (\emph{a - e}) ARPES intensity plots of the band structure at $\Gamma$ at different temperatures. The intensity in each plot is the sum of data acquired with C$^+$ and C$^-$ polarized photons. (\emph{f - j}) EDC curvatures of (\emph{a - e}). (\emph{k - o}) ARPES intensity plots of the band structure at the M point at different temperatures, recorded in the $\sigma$ geometry.  (\emph{p - t}) MDC curvatures of (\emph{k - o}). (\emph{u}) EDCs of (\emph{a - e}) at $k_x =0$. The blue dashed line is the second derivative of the blue solid line with an extra minus sign. The two dashed black lines indicate the peak positions. (\emph{v}) EDCs of (\emph{k - o}) at $k_x =0$. The black dashed lines correspond to the EDC peaks. In all the intensity plots, the red lines represent the d$_{yz}$ orbitals, while the blue/cyan ones represent the d$_{xz}$ orbitals. All cuts are along $\Gamma$-M or the Z-A high-symmetry direction. All the intensities are divided by Fermi function at the corresponding temperatures. }
\end{figure*}


We show in Fig. \ref{band} the electronic band structure of FeSe, below the structural transition. The Fermi surface (FS) (Fig. \ref{band}(\emph{a})) is formed by one hole pocket centered at Z and two electron pockets centered at A. Based on local density approximation (LDA) calculations (Supp. Part I), we attribute the two elliptical electron pockets at A to $d_{xz/yz}$ bands in different twin domains, while the $d_{xy}$ electron pocket is not observed. A schematic representation of the FSs ($T > T_s$) and their areas are shown in Supp. Part III. As shown in Fig. \ref{band}(\emph{b}), the $k_z$ dispersion along $\Gamma$-Z is non-negligible, in agreement with a previous report \cite{BorisenkoPRB2014}. The Fermi wave vector ($k_F$) near the $\Gamma$ point is $\sim 0.07\pi/a$, while it is $\sim 0.14\pi/a$ at the Z point. The small $k_F$ can be clearly resolved from the cut at $\Gamma$ displayed in Figs. \ref{band}(\emph{c}), \ref{band}(\emph{f}) and \ref{band}(\emph{i}). Besides the $d_{yz}$ band, we also resolve a steep $d_{xz}$ and a flat $d_{xy}$ bands below $E_F$. Interestingly, the top of the $d_{xz}$ and $d_{yz}$ bands do to coincide, in contrast to LDA calculations but in agreement with a previous ARPES report \cite{Coldeaarxiv2015}. From the energy distribution curves (EDCs) and curvature intensity plots \cite{curvature}, we estimate that the splittings at $\Gamma$ is about 30 meV at T $\sim$ 20 K. As shown in Fig. \ref{band}(f), we notice that there is a large hybridization gap between the $d_{yz}$ and $d_{xy}$ bands near $\Gamma$ but little hybridization or none between the $d_{xy}$ and $d_{xz}$ bands. 

The band structure at Z (Figs. \ref{band}(\emph{d}), \ref{band}(\emph{g}) and \ref{band}(\emph{j})) is very similar, except for a relative shift along the energy direction. In particular, a splitting of about 30 meV is observed at Z between the $d_{xz}$ and $d_{yz}$ bands, and an hybridization gap is found between the $d_{xy}$ and $d_{yz}$ bands, but not between the $d_{xy}$ and $d_{xz}$. In Figs. \ref{band}(\emph{e}), \ref{band}(\emph{h}) and \ref{band}(\emph{k}), we show the band structure at M. We distinguish two hole-like bands associated with the $d_{yz}$ bands from different twin domains. Because of a lack of coherence, the $d_{xy}$ electron and hole bands at M are not observed. Our data indicate that the splitting at M is about 50 meV at T $\sim$ 50 K, which is quite different from the prediction of onsite interactions.

To fully understand the splittings and check if they are related, we performed temperature-dependent experiments. The temperature evolution of the $d_{xz}/d_{yz}$ splittings at high-symmetry points is illustrated in Fig. \ref{split}. Except for thermal broadening, the intensity plots show that the band dispersions around $\Gamma$ barely change with temperature and that the separation between the $d_{xz}$ and $d_{yz}$ bands is nearly temperature independent. In other words, the $d_{xz}/d_{yz}$ splitting at $\Gamma$ is almost not changed within the temperature range studied, and the hybridization gap between the $d_{xy}$ band and the $d_{yz}$ band persists at high temperature, whereas no hybridization is found between the $d_{xy}$ band and the $d_{xz}$
band, indicating that none of these phenomena is directly related to the structural transition. Our conclusion on the splitting at $\Gamma$ is reinforced by the comparison of the EDCs at the $\Gamma$ point, displayed in Fig. \ref{split} (\emph{u}), and at the Z point (see Supp. FigS2). 

Unlike our observation at $\Gamma$/Z, the band splitting at M varies strongly with temperature. The two sets of bands from different domains gradually merge with increasing temperature. At $T = 120$ K, we only see one set of band structure, which implies the disappearance of the domain structure, in agreement with previous results \cite{TakahashiPRL2014, ShimojimaPRB2014, Coldeaarxiv2015}. The evolution of the EDCs with temperature at M is shown in Fig. \ref{split} (\emph{v}). The dashed lines mark the two sets of band tops/bottoms merging at $T = 120$ K.

\begin{figure}[!t]
\begin{center}
\includegraphics[width=0.48\textwidth]{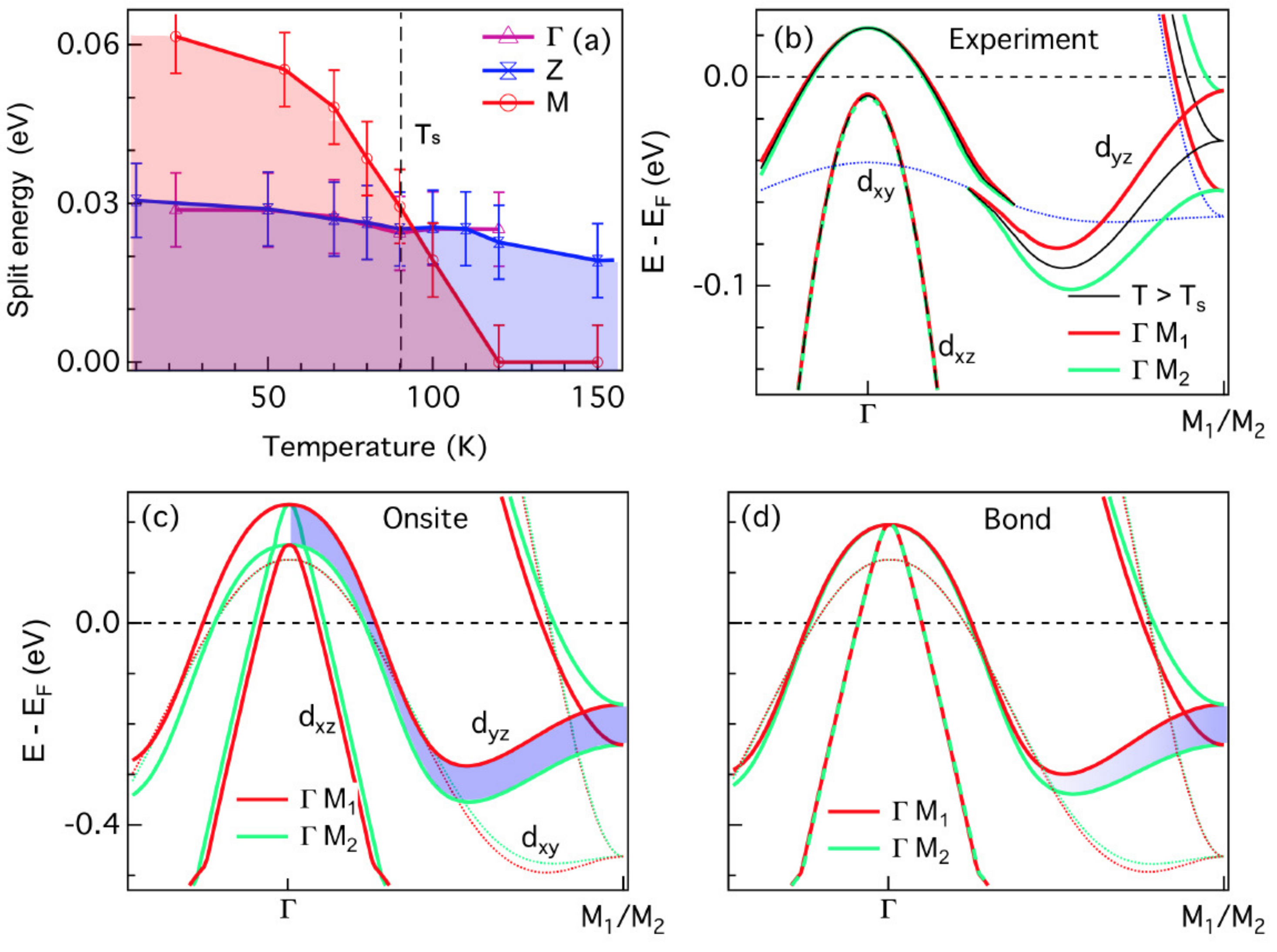}
\end{center}
 \caption{\label{cartoon} \emph{(a)} Summary of the $d_{xz}/d_{yz}$ splittings at $\Gamma$, Z and M as a function of temperature. The splitting at M disappears at a slightly higher temperature than $T_s$, which might be caused by short-range ordering or fluctuations above the transition. \emph{(b)} Draft of band structure extracted from experimental data and fitted with a tight binding model. The $d_{xy}$, $d_{xz}$ and $d_{yz}$ symbols are for $T> T_s$ only since the orbital characters will change along $\Gamma$-M$_1$ and $\Gamma$-M$_2$. \emph{(c - d)} Band structure of onsite and bond orbital order, calculated from a tight binding model (see Supp. Part I for the details). }
\end{figure}

Fig. \ref{cartoon}(\emph{a}) compares the temperature dependence of the different splittings and Fig. \ref{cartoon}(\emph{b}) gives a schematic representation of the experimental splittings and hybridizations observed below and above $T_s$. The splittings at $\Gamma$ and Z have the same amplitude, which varies very slowly with temperature, even across the structural transition. In sharp contrast, the splitting at the M point is nearly twice that at the $\Gamma$ point at low temperature, but it decreases with temperature and vanishes at 100 - 120 K. We conclude that we must introduce two parameters to explain the data. The splitting at $\Gamma$/Z is temperature independent and affects only the BZ area around $\Gamma$ and Z, while the parameter inducing the splitting at M only affects the M point and is related to structural transition. Since it does not affect the splitting at the BZ center, the order parameter responsible for the splitting at the M point must have an anisotropic form of orbital order, such as the $d$-wave orbital order defined on the Fe-Fe bonds \cite{LiJPCM2015}: 

\begin{eqnarray*}
&& H_{bond} =\sum_{\vec k}\Delta_{\textrm{M}}(T) (\cos k_x-\cos k_y)(n_{xz}(\vec k)+n_{yz}(\vec k)),
\end{eqnarray*}

In Fig.\ref{cartoon}(\emph{d}), we provide detailed calculations and show that the $d$-wave orbital order can explain the experimental band structure near the M point very well with an estimated coupling constant $\Delta_0\sim 60 $ meV in the low-temperature limit.

Two major candidates for the splitting at $\Gamma$/Z are the SOC \cite{FernandesPRB2014} and the onsite ferro-orbital fluctuations \cite{LiJPCM2015}. However, both explanations contain severe flaws. Indeed, SOC can break the glide symmetry that prevents the $d_{xy}$ band at $k+Q$ to hybridize with the $d_{xz}/d_{yz}$ bands at $k$  in the 1-Fe unit cell (Supp. Part I) \cite{HirschfeldRP2011, HuPRX2013}. However, such hybridization has an equal strength for both $d_{xy,\uparrow}/d_{xz,\downarrow}$ and $d_{xy,\uparrow}$/ $d_{yz,\downarrow}$ hybridizations. Thus, the observation of hybridization between the $d_{yz}$ and $d_{xy}$ bands but not between the $d_{yz}$ and $d_{xy}$ bands is strongly against the SOC origin \footnote{We note that the broken $z$-mirror symmetry at the surface may also cause such hybridization. However, the interlayer coupling is believed to be small in FeSe, which is unlikely to cause the strong effect observed here.}. In addition, similar splitting at $\Gamma$ has been reported to be strongly doping dependent in LiFeAs, which is apparently in contradiction with the SOC scenario \cite{DingPRB2014}. The other candidate, the onsite ferro-orbital ordering or fluctuations, should remove the $d_{xz}/d_{yz}$ degeneracy across the entire momentum space, as illustrated by our calculations shown in Fig.\ref{cartoon}(\emph{c}), which is inconsistent with the absence of splitting at M above 120 K. Together with the doping-dependent splitting observed in LiFeAs \cite{DingPRB2014}, we have strong reasons to believe that in FeSe the splitting at $\Gamma$ and the hybridization between the $d_{xy}$ and $d_{yz}$ bands originate from magnetic fluctuations. The magnetism in FeSe is more frustrated than in the iron-pnictides, and long-range magnetic ordering is thus unstable \cite{DaiNC2011, HuPRB2012}. However, nematicity and magnetic fluctuations can still be strongly coupled  \cite{Mazinarxiv2015, Siarxiv2015}. Thus, the splitting at $\Gamma$ and the hybridization between $d_{xy}$ and $d_{yz}$ observed above $T_s$ are very likely signatures of this coupling. In any cases, our current results with two distinct $d_{xz}/d_{yz}$ splittings suggest a more complicated interplay between the magnetic and orbital degrees in FeSe than previously expected. 


In conclusion, we report the temperature evolution of the detailed electronic band structure in FeSe single-crystals. We observe two distinct $d_{xz}/d_{yz}$ band splittings at the high-symmetry points. The splitting at M is related to the structural transition and has a $d$-wave form factor, while the splitting at $\Gamma$ originates most likely from magnetic frustration. Our results clearly exclude the commonly-believed ferro-orbital order and require a new consideration of the origin and implication of the orbital order in FeSCs.

We acknowledge D. H. Lee, T. Li and K. Jiang for useful discussions. This work was supported by grants from CAS (XDB07000000), MOST (2015CB921300, 2011CBA001000, 2013CB921700, 2012CB821400), NSFC (11474340, 11274362, 11234014, 11190020, 91221303, 11334012) and US DOE BES grants DE-SC0002554 and DE-FG02-99ER45747.




\bibliographystyle{apsrev}
\bibliography{biblio_FeSe}

\end{document}